\documentclass[12pt]{article}
\usepackage{amsmath,lscape,amsfonts,amssymb,amsthm,verbatim,color,lscape}
\usepackage[figuresright]{rotating}
\usepackage[sort&compress]{natbib}
\usepackage{url}
\usepackage{hyperref}
\hypersetup{colorlinks,
citecolor=black,
filecolor=black,
linkcolor=black,
urlcolor=black,
pdftex}

\usepackage{times}
\usepackage{blindtext}

\usepackage{sectsty}  
\subsectionfont{\normalsize\bf}
\sectionfont{\large\bf}

\begin{document}
\def\a{\alpha}
\def\b{\beta}
\def\g{\gamma}
\newcommand{\bea}{\begin{eqnarray}}
\newcommand{\eea}{\end{eqnarray}}
\def\s{\sigma}
\def\R{\mathbf{R}}
\def\D{\mathbf{D}}
\def\A{\mathbf{A}}
\def\S{\boldsymbol{\Sigma}}
\def\I{\mathbf{I}}
\def\x{\mathbf{x}}
\def\X{\mathbf{X}}
\def\bfv{\mathbf{v}}
\def\V{\mathbf{V}}
\def\Y{\mathbf{Y}}
\def\y{\mathbf{y}}
\def\bbf{\boldsymbol{\beta}}
\def\gbf{\boldsymbol{\gamma}}
\def\obf{\boldsymbol{\omega}}
\def\Obf{\boldsymbol{\Omega}}
\def\U{\mathbf{U}}
\def\J{\mathbf{J}}
\def\ga{\gamma}
\def\u{\mathbf{u}}
\def\C{\mathbf{C}}
\def\p{\mathbf{p}}
\def\1{\mathbf{1}}
\def\t{\mathbf{t}}
\def\W{\mathbf{W}}
\def\thbf{\boldsymbol{\theta}}

\title{Efficient estimation of  high-dimensional\\ multivariate normal copula  models\\ with discrete spatial responses}

\date{}
\author{
Aristidis K. Nikoloulopoulos\footnote{{\small\texttt{A.Nikoloulopoulos@uea.ac.uk}}, School of Computing Sciences, University of East Anglia,
Norwich NR4 7TJ, UK}
}
\maketitle

\vspace{2ex}

\begin{abstract}
\baselineskip=24pt
\noindent
The distributional transform (DT) is amongst the computational methods used for estimation of high-dimensional multivariate normal 
copula models with discrete responses. Its advantage is that the likelihood can  be derived conveniently under the theory for copula models with continuous margins, but there has not been a clear analysis of the adequacy of this method. We investigate the small-sample  and asymptotic efficiency of the method  for estimating high-dimensional multivariate normal copula models with univariate Bernoulli, Poisson, and negative binomial margins, and  show that the  DT approximation leads to biased estimates when there is more discretization.
For a high-dimensional discrete response, we implement  a maximum simulated likelihood  method, which is based  on evaluating the multidimensional integrals of the likelihood with  randomized quasi Monte Carlo methods. Efficiency calculations show that our method is nearly as efficient as maximum likelihood for fully specified high-dimensional multivariate normal 
copula models. 
Both methods are illustrated with spatially aggregated count data sets, and it is shown that there is a substantial gain on efficiency via the maximum simulated likelihood method. 
\end{abstract}

\maketitle

\baselineskip=24pt

\section{Introduction}
There are classical statistical models in the literature for regression and prediction with spatial continuous data for which parameter estimation and inference are straightforward. These models are  based on the multivariate normal (MVN) distribution, applied to a possibly transformed response (for a thorough review see \cite{diggle-Ribeiro-2007}). But applications of these models are invalid for discrete spatial categorical and count response data. These data  occur in several disciplines such as epidemiology, ecology, agriculture, to
name just a few. Flexible models for such data are not widely available and usually hard to fit due to the fact that  available multivariate discrete distributions can have only certain properties, as for example they provide limited dependence or
they can have marginal distributions of a given form.  Most of the existing literature is concerned with the generalized linear mixed model proposed by \cite{diggle-etal-1998}. This model was initially proposed for the analysis of correlated count data by \cite{aitchinson&ho89}. Limitations of this model are that the marginal distributions belong to a specific parametric family, and,  because
the correlation structure come from a continuous multivariate mixing
distribution, the possible choices are very limited.
Thus, it seems that there is a lack of flexible models appropriate for spatial discrete data or flexible marginal choices.

In this paper, we use copulas (distributions with uniform margins on the unit interval) to overcome this problem. The power of copulas for dependence modelling is due to the dependence structure being considered separate from the univariate margins; see,
for example, Section 1.6 of \cite{joe97}.
A discrete regression
spatial model through copulas is not new. A copula approach was recently proposed and studied by \cite{madsen09}, \cite{Kazianka&Pilz2010}, \cite{Kazianka2013} and \cite{Hughes2014} who explored the use of MVN copulas to describe the distribution of  geostatistical and spatially aggregated  data, respectively.

Our
empirical experience is that
the MVN copula model
with discrete margins
provides the best or nearly the best fit. 
However, implementation of the MVN copula for discrete data is possible,
but not easy, because the MVN distribution as a latent model for discrete response requires rectangle probabilities based on multidimensional integrations \citep{nikoloulopoulos&joe&chaganty10}. \cite{song07}, influencing other authors (e.g.  \cite{Kazianka2013}, \cite{Hughes2014}), acknowledged  that the probability mass function (pmf) can be obtained  as a finite difference of the copula cumulative distribution function (cdf). Generally speaking, this is an imprecise statement, since calculating the finite difference among $2^d$ (where $d$ is the dimension) numerically computed orthant probabilities  may result
in negative values. The pmf can be alternatively obtained by computing  a MVN rectangle probability.
The randomized quasi Monte Carlo methods proposed by \cite{genz92} and \cite{genz&bretz02} can be used for that purpose.
If one computes the rectangle MVN probabilities via simulation based on the methods in \cite{genz92} and \cite{genz&bretz02}, then one is
using a simulated likelihood method. \cite{nikoloulopoulos13b} has studied the asymptotic and small-sample efficiency of  this simulated likelihood method  and has shown it is  as
good as maximum likelihood for dimension 10 or lower.

\cite{madsen09} and \cite{Madsen&Fang11} proposed a different simulated likelihood method to ``approximate" the likelihood, by using the continuous extension (CE) of a discrete random variable developed in \cite{denuit&lambert05} and applied the method to geostatistical and longitudinal discrete data, respectively. \cite{Shi&valdez2014a,Shi&valdez2014b} also used the simulated likelihood method based on the CE for longitudinal and multivariate insurance claim counts. 

\cite{Kazianka&Pilz2010} and \cite{Kazianka2013} proposed and studied respectively a fast  surrogate likelihood method, by approximating the rectangle probability with a copula density using the distributional transform (DT); see  \cite{Ferguson-1967} for an early appearance. Both the simulated likelihood method based on the CE and the surrogate likelihood method based on the DT  have the advantage that the likelihood can  be derived conveniently under the theory for copula models with continuous margins.  

\cite{Hughes2014} recommended the simulated likelihood method based on the CE and the surrogate likelihood method based on the DT along with the composite likelihood (CL) method \citep{varin08} for modelling  spatially aggregated discrete (areal) data and  used the  simulated likelihood method based on the CE  to ``judge" the DT and CL approaches. It came as somewhat of a surprise to us, because (a) asymptotic and small-sample efficiency calculations in \cite{nikoloulopoulos13b} have  shown that the  simulated likelihood method based on the CE   is a very inefficient approach;  it leads  to substantial downward bias for the estimates  of the latent correlation and the univariate marginal parameters that they are not regression coefficients for fully specified multivariate normal copula-based models, and, (b) CL methods are well established as an alternative of maximum likelihood  when the joint probability is too difficult to compute; they lead to unbiased estimating equations \citep{Varin-etal2011}. Note in passing that they have been studied for copula modelling  by \cite{zhao&joe05}.

The surrogate likelihood method based on the DT cannot be recommended until its properties have been studied and compared to efficient existing  methods.
The DT method  has been previously used in copula literature, e.g. to prove stochastic ordering results and construct tests of dependence properties in \cite{Ruschendorf1981} and \cite{Ruschendorf09}, respectively.
Although its application to copula dependence modelling for discrete data  is novel, its asymptotic properties  has yet to be established in that context.

The contribution in this paper is (a) to  examine thoroughly the accuracy and the adequacy of the surrogate likelihood method based on the DT using asymptotics and small-sample efficiency studies;  (b) to study the estimation of the MVN copula model with discrete responses via the simulated likelihood method proposed by \cite{nikoloulopoulos13b} in a high-dimensional/spatial context.

The remainder of the paper proceeds as follows. Section \ref{sec-model} has a brief overview for  MVN copula models.
Likelihood estimation methods are provided in Section \ref{sec-estimation}.
Section \ref{sec-asym} and Section \ref{sec-sim} contain  theoretical  (asymptotic properties of the estimators) and small-sample  efficiency calculations, respectively,
to assess the accuracy of the discussed  likelihood  estimation methods.
Section \ref{sec-app} 
 presents applications of the likelihood estimation methods to two  areal data sets. In these examples, it turns out that the  surrogate likelihood  method based on the DT  could lead to invalid inference. 
We conclude with some discussion in Section \ref{sec-disc}.

\section{\label{sec-model}Overview and relevant background for MVN copula models}

A copula is a multivariate cdf with uniform $U(0,1)$ margins \citep{joe97,nelsen06,joe2014}.
If $F$ is a $d$-variate cdf with univariate margins $F_1,\ldots,F_d$,
then Sklar's (1959) theorem\nocite{sklar1959} implies that there is a copula $C$ such that
  $$F(y_1,\ldots,y_d)= C\Bigl(F_1(y_1),\ldots,F_d(y_d)\Bigr).$$
The copula is unique if $F_1,\ldots,F_d$ are continuous, but not
if some of the $F_j$ have discrete components.
If $F$ is continuous and $(Y_1,\ldots,Y_d)\sim F$, then the unique copula
is the distribution of $(U_1,\ldots,U_d)=\left(F_1(Y_1),\ldots,F_d(Y_d)\right)$ leading to
\begin{equation}\label{inversion-method}
C(u_1,\ldots,u_d)=F\Bigl(F_1^{-1}(u_1),\ldots,F_d^{-1}(u_d)\Bigr),
  \quad 0\le u_j\le 1, j=1,\ldots,d,
\end{equation}
where $F_j^{-1}$ are inverse cdfs. In particular,
if $\Phi_d(\cdot;\R)$
is the MVN cdf with correlation matrix $$\R=(\rho_{jk}: 1\le j<k\le d)$$ and
N(0,1) margins, and $\Phi$ is the univariate standard normal cdf,
then the MVN copula is
$$C(u_1,\ldots,u_d;\R)=\Phi_d\Bigl(\Phi^{-1}(u_1),\ldots,\Phi^{-1}(u_d);\R\Bigr).$$

Consider a multivariate discrete regression setup in which $d \geq 2$ dependent discrete random variables $Y_{1}, \ldots, Y_{d}$ are observed together with a vector $\mathbf{x} \in \mathbb{R}^p$ of explanatory variables.
If $C(\cdot;\R)$ is the MVN copula (or any other  parametric
family of copulas) and $F_j(\cdot\; ;\; \nu,\gbf)$,
where $\nu=\eta(\x;\bbf)$ is a function of $\x$
and the $p$-dimensional regression vector $\bbf$, and $\gbf$ is the $r$-dimensional  vector of univariate parameters that are not regression coefficients, 
is a  parametric model for the $j$th univariate margin 
then
  $$C\Bigl(F_1(y_1;\nu_1,\gbf),\ldots,F_d(y_d;\nu_d,\gbf);\R\Bigr)$$
is a multivariate parametric model with univariate margins $F_1,\ldots,F_d$.
For copula models, the response vector $\Y=(Y_1,\ldots,Y_d)$ can be  discrete \citep{Nikoloulopoulos2013a,nikoloulopoulos&joe12}.

\section{\label{sec-estimation}Likelihood estimation methods}
In this section,
we discuss the simulated likelihood method proposed by \cite{nikoloulopoulos13b} and the  surrogate likelihood method  based on the DT \citep{Kazianka&Pilz2010,Kazianka2013,Hughes2014}. The simulated likelihood method based on the CE \citep{madsen09,Madsen&Fang11,Shi&valdez2014a,Shi&valdez2014b,Hughes2014} is not discussed/used in the sequel since its inefficiency has been shown  in \cite{nikoloulopoulos13b} and should be avoided for copula  dependence modelling with multivariate/longitudinal discrete data.   
\subsection{Simulated likelihood}
For a sample of size $n$ with data $\y_1,\ldots,\y_n$,
the joint log-likelihood of a MVN copula model is
\begin{equation}\label{MLlik}
\ell(\bbf,\gbf,\R)\\
= \sum_{i=1}^{n}
\log{h(y_{i1},\ldots,y_{id};\bbf,\gbf,\R)},
\end{equation}
where $h(\cdot;\bbf,\gbf,\R)$ is the
joint pmf of the multivariate discrete response vector  $\Y$.
The pmf can be obtained by computing  the following rectangle probability,
\begin{eqnarray}\label{MVNpmf}
h(\y;\bbf,\gbf,\R)&=&\Pr (Y_1 = y_1, \ldots , Y_d = y_d;\mathbf{x})\\
&=&\Pr(y_1-1< Y_1\leq y_1,\ldots,y_d-1< Y_d\leq y_d;\mathbf{x})\nonumber\\
&=&\int_{\Phi^{-1}[F_{1}(y_1-1;\nu_1,\gbf)]}^{\Phi^{-1}[F_{1}(y_1;\nu_1,\gbf)]}\cdots
\int_{\Phi^{-1}[F_{d}(y_d-1;\nu_d,\gbf)]}^{\Phi^{-1}[F_{d}(y_d;\nu_d,\gbf)]}  \phi_\R(z_1,\ldots,z_d) dz_1\ldots dz_d,\nonumber
\end{eqnarray}
where $\phi_\R$ denotes the standard MVN density  with latent correlation
matrix $\R$.

There are several papers in the literature that focus on the
computation of the MVN rectangle probabilities.  The dominant of the methods is a quasi Monte Carlo method proposed by \cite{genz92} and \cite{genz&bretz02}. The method  achieves error reduction of Monte Carlo methods with variance reduction methods as (a) transforming to a bounded integrand, (b) using antithetic variates, and (c) using a randomized quasi Monte Carlo method. The test results in \cite{genz&bretz02,genz&bretz2009} show that
the method is very efficient, compared to other methods in the literature.
Note in passing that the implementation of the proposed algorithms in \cite{genz&bretz02} is available in the {\it mvtnorm} package in R \citep{genz-etal-2012}. This advance in computation of MVN probabilities can be used to implement high-dimensional MVN copula models with discrete response data.

 \cite{nikoloulopoulos13b} proposed  a simulated likelihood (hereafter SL) method, where the rectangle MVN probabilities in (\ref{MLlik}) are computed based on the methods in \cite{genz&bretz02}. The estimated parameters can be obtained by  maximizing  the simulated log-likelihood in (\ref{MLlik})
over the univariate and copula parameters $(\bbf,\gbf,\R)$. Since the estimation of the parameters of the MVN copula-based models
is obtained using a quasi-Newton routine \citep{nash90} applied to the log-likelihood in (\ref{MLlik}),  the use
of randomized quasi Monte Carlo simulation to  four decimal place
accuracy for evaluations of integrals works poorly, because
numerical derivatives of the log-likelihood with respect to
the parameters are not smooth. In order to achieve smoothness, the same set of uniform random variables should be used for every rectangle probability that comes up in the optimization of the SL. The method was initially proposed for the analysis of discrete longitudinal  data \citep{nikoloulopoulos13b}. We refer the interested reader to this paper for more details.

\subsection{Surrogate likelihood based on the DT}

Copula models were originally developed for continuous responses where the density is obtained using partial derivatives of the multivariate copula cdf,  and hence the numerical calculations are much simpler.
\cite{Kazianka&Pilz2010} and \cite{Kazianka2013} proposed and studied respectively a surrogate likelihood method, by approximating the rectangle probability in (\ref{MVNpmf}) with a copula density using the DT  \citep{Ferguson-1967}.

The surrogate likelihood takes the form
\begin{equation}\label{kalik}
\ell(\bbf,\gbf,\R)\\
\approx \sum_{i=1}^{n}
\log{c(v_{i1},\ldots,v_{id};\R)}+\sum_{i=1}^{n}\sum_{j=1}^{d} \log f_j(y_{ij};\nu_{ij},\gbf),
\end{equation}
where $v_j=0.5\Bigl(F_j(y_j;\nu_j,\gbf)+F_j(y_j-1;\nu_j,\gbf)\Bigl)$ and $f_1(y_1;\nu_1,\gbf),\ldots, f_d(y_d;\nu_d,\gbf)$ are the univariate marginal pmfs. 
Since the MVN copula has a closed form density
$$c(u_1,\ldots,u_d;\R)=|\R|^{-1/2}\exp\Bigl[\frac{1}{2}\bigl\{\mathbf q^\top(\I_d-\R^{-1})\,\mathbf q\bigr\}\Bigr],$$
where $\mathbf q=(q_1,\ldots,q_d)$ with $q_j=\Phi^{-1}(u_j), j=1,\ldots,d$ and $\I_d$ is the $d$-dimensional identity matrix,
the authors avoid the  multidimensional  integration, and hence the numerical calculations are much simpler and faster.
The estimated parameters can be obtained by  maximizing  the surrogate log-likelihood in (\ref{kalik})
over the univariate and copula parameters $(\bbf,\gbf,\R)$.

\section{\label{sec-asym}Asymptotics}
In this section, we study the asymptotics of  the surrogate likelihood method based on the DT \citep{Kazianka&Pilz2010,Kazianka2013,Hughes2014}, along with the asymptotics of the SL in \cite{nikoloulopoulos13b}, and we assess the accuracy based on the limit (as the number of clusters
increases to infinity) of the maximum surrogate likelihood estimate  (DTMLE) and the maximum SL estimate  (MSLE). We restrict ourselves to a MVN copula model with a positive  exchangeable  dependence structure, i.e., we took
$\R$ as $(1-\rho)\I_d+ \rho \J_d$,
where  
$\J_d$ is the
$d\times d$ matrix of 1s. For positive exchangeable correlation structures, the $d$-dimensional integrals
conveniently reduce to 1-dimensional integrals \citep[p. 48]{Johnson&Kotz72}. Hence,
MVN  rectangle probabilities can be quickly
computed to a desired accuracy that is $10^{-6}$ or less, because
1-dimensional numerical integrals are computationally easier than
higher-dimensional numerical integrals. If one computes the rectangle MVN probabilities  in (\ref{MLlik}) with the 1-dimensional integral method in \cite{Johnson&Kotz72}, then one is using a numerically accurate likelihood method that is valid for any dimension \citep{nikoloulopoulos13b}.

By varying factors such as dimension $d$,  the amount of discreteness (binary versus count response), and latent correlation  for exchangeable structures,   we demonstrate patterns
in the asymptotic bias of the DTMLE and MSLE, and assess the performance of the surrogate and simulated likelihood.
Note that the performance of the SL method has been already assessed in \cite{nikoloulopoulos13b} and it is shown that it is good as maximum likelihood for dimension 10 or lower.
For the cases where we compute the probability limit, we will take a constant dimension $d$ that increases. For marginal models we use Bernoulli$(\mu)$, Poisson$(\mu)$, and  negative binomial (NB). For the latter model, we use both the NB$1(\mu,\,\ga)$  and NB$2(\mu,\,\ga)$ parametrization in \cite{cameron&trivedi98}; the NB$2$ parametrization is that used in \cite{lawless87}.
For ease of exposition, we also consider
the case that $\mu$ is common to different univariate margins and does not depend on covariates.

Let the $T$ distinct cases for the discrete response be denoted as
$$\y^{(1)},\ldots, \y^{(T)},\quad \y^{(t)}=(y_1^{(t)},\ldots,y_d^{(t)}),\, t=1,\ldots, T.$$
In a random sample of size $n$, let the corresponding frequencies be denoted as $n^{(1)},\ldots, n^{(T)}$ and  $p^{(t)}$ be the limit in probability of $n^{(t)}/n$ as $n\to \infty$.
For the SL in (\ref{MLlik}), we have the limit
\begin{equation}\label{limitslik}
n^{-1}\ell(\mu,\ga,\rho)\to \sum_{t=1}^{T} p^{(t)}
\log{h(y_{1}^{(t)},\ldots,y_{d}^{(t)};\mu,\ga,\R)},
\end{equation}
where $h(\mathbf y^{(t)};\mu,\ga,\R)$ is computed using the method in \cite{genz&bretz02}.
The limit of the  MSLE (as $n\to \infty$) is the maximum of (\ref{limitslik});
we denote this limit as $(\mu^{SL},\ga^{SL},\rho^{SL})$. Note in passing that the limit of the  standard MLE (as $n\to \infty$) is the maximum of (\ref{limitslik}) where $h(\mathbf y^{(t)};\mu,\ga,\R)$ is computed  with the 1-dimensional integral method in \cite{Johnson&Kotz72}.
For the surrogate log-likelihood in (\ref{kalik}), we have the limit
\begin{equation}\label{limitlik}
n^{-1}\ell(\mu,\ga,\rho)\to \sum_{t=1}^{T} p^{(t)}\Bigl\{
\log{c(v_{1}^{(t)},\ldots,v_{d}^{(t)};\R)}+\sum_{j=1}^{d} \log f_j(y_j^{(t)};\mu,\ga)\Bigr\},
\end{equation}
where $v^{(t)}=0.5\Bigl(F_j(y^{(t)};\mu,\ga)+F_j(y^{(t)}-1;\mu,\ga)\Bigl)$.
The limit of the  DTMLE (as $n\to \infty$) is the maximum of (\ref{limitlik});
we denote this limit as $(\mu^{DT},\ga^{DT},\rho^{DT})$.

We will compute these limiting  MSLE  and DTMLE  in a variety of situations to show clearly if  the simulated and surrogate likelihood methods
are good. By using these limits, we do not need Monte Carlo simulations for comparisons, and we can quickly vary parameter
values and see the effects.
The $p^{(t)}$
in (\ref{limitslik}) and (\ref{limitlik}) are the model based probabilities $h(\mathbf y^{(t)};\mu,\ga,\R)$, and computed with the 1-dimensional integral method in \cite{Johnson&Kotz72}.  For a count response, we get a finite number of $\y^{(t)}$ vectors by truncation. The truncation point  is chosen to exceed $0.999$ for total probabilities.

\begin{table}[!h]
\begin{center}

\begin{tabular}{ccccc|ccccc}
\hline
    $\mu$ &           \multicolumn{ 4}{c|}{$\mu^{{DT}}$} &    $\rho$ &          \multicolumn{ 4}{c}{$\rho^{{DT}}$} \\

           &     $d=2$ &     $d=3$ &     $d=5$ &    $d=10$ &            &     $d=2$ &     $d=3$ &     $d=5$ &    $d=10$ \\
\hline
       0.2 &      0.225 &      0.245 &      0.269 &      0.290 &        0.2 &      0.605 &      0.643 &      0.674 &      0.696 \\

       0.5 &      0.500 &      0.500 &      0.500 &      0.500 &        0.2 &      0.488 &      0.523 &      0.555 &      0.579 \\

       0.8 &      0.775 &      0.755 &      0.731 &      0.710 &        0.2 &      0.605 &      0.643 &      0.674 &      0.696 \\

       0.2 &      0.232 &      0.253 &      0.274 &      0.293 &        0.5 &      0.715 &      0.736 &      0.752 &      0.764 \\

       0.5 &      0.500 &      0.500 &      0.500 &      0.500 &        0.5 &      0.650 &      0.664 &      0.677 &      0.686 \\

       0.8 &      0.768 &      0.747 &      0.726 &      0.707 &        0.5 &      0.715 &      0.736 &      0.752 &      0.764 \\

       0.2 &      0.240 &      0.261 &      0.280 &      0.297 &        0.8 &      0.834 &      0.842 &      0.849 &      0.854 \\

       0.5 &      0.500 &      0.500 &      0.500 &      0.500 &        0.8 &      0.800 &      0.805 &      0.808 &      0.811 \\

       0.8 &      0.760 &      0.739 &      0.720 &      0.703 &        0.8 &      0.834 &      0.842 &      0.849 &      0.854 \\
\hline
\end{tabular}
\caption{\label{be-bias} Limiting DTMLE for MVN copula-based models with Bernoulli$(\mu)$ margins.}
\end{center}
\end{table}

Representative results are shown in Tables \ref{be-bias} and \ref{nb2-bias} for Bernoulli$(\mu)$, and  NB2$(\mu,\,\ga)$ margins, with MSLE results omitted because they were identical with MLE
up to three or four decimal places. Therefore, the SL method leads to unbiased estimates.
As regard as the surrogate likelihood method, for binary responses there is substantial asymptotic bias for both the univariate and latent correlation parameters. The only case that there is asymptotic  unbiasedness for $\mu$ is when $\mu=0.5$. This is due the fact that the individual probabilities have the same size. For non-overdispersed count responses the surrogate likelihood method leads to approximate asymptotic unbiasedness for both the univariate and latent correlation parameters as $\mu$ increases. However, for overdispersed count responses, the method leads to asymptotic bias (decreases as $\mu$ increases or $\ga$ or $\rho$ or $d$ decreases) for all the univariate parameters and substantial asymptotic upward bias for the latent correlation $\rho $ (decreases as $\rho$ increases).

\begin{table}[!h]
\begin{center}

\begin{tabular}{ccc|ccc|ccc}
\hline
    $\mu$ & \multicolumn{ 2}{c|}{$\mu^{{DT}}$} &  $\gamma$ & \multicolumn{ 2}{c|}{$\gamma^{{DT}}$} &    $\rho$ & \multicolumn{ 2}{c}{$\rho^{{DT}}$} \\

           &     $d=2$ &     $d=3$ &            &     $d=2$ &     $d=3$ &            &     $d=2$ &     $d=3$ \\
\hline
       0.5 &      0.504 &      0.526 &        0.5 &      0.603 &      0.759 &        0.2 &      0.348 &      0.419 \\

         1 &      0.994 &      0.992 &        0.5 &      0.530 &      0.567 &        0.2 &      0.259 &      0.283 \\

         2 &      1.992 &      1.985 &        0.5 &      0.510 &      0.521 &        0.2 &      0.225 &      0.234 \\

         5 &      4.993 &      4.987 &        0.5 &      0.502 &      0.505 &        0.2 &      0.208 &      0.210 \\

        10 &      9.995 &      9.991 &        0.5 &      0.501 &      0.501 &        0.2 &      0.203 &      0.204 \\

       0.5 &      0.522 &      0.604 &          2 &      2.248 &      2.632 &        0.2 &      0.390 &      0.498 \\

         1 &      1.001 &      1.020 &          2 &      2.103 &      2.232 &        0.2 &      0.292 &      0.335 \\

         2 &      1.989 &      1.987 &          2 &      2.055 &      2.118 &        0.2 &      0.252 &      0.272 \\

         5 &      4.976 &      4.956 &          2 &      2.029 &      2.059 &        0.2 &      0.228 &      0.238 \\

        10 &      9.964 &      9.932 &          2 &      2.018 &      2.037 &        0.2 &      0.219 &      0.225 \\

       0.5 &      0.534 &      0.594 &        0.5 &      0.739 &      0.959 &        0.5 &      0.618 &      0.664 \\

         1 &      1.010 &      1.035 &        0.5 &      0.601 &      0.693 &        0.5 &      0.565 &      0.594 \\

         2 &      1.997 &      2.002 &        0.5 &      0.541 &      0.581 &        0.5 &      0.535 &      0.552 \\

         5 &      4.993 &      4.989 &        0.5 &      0.511 &      0.521 &        0.5 &      0.513 &      0.520 \\

        10 &      9.994 &      9.990 &        0.5 &      0.504 &      0.507 &        0.5 &      0.505 &      0.508 \\

       0.5 &      0.584 &      0.752 &          2 &      2.467 &      2.861 &        0.5 &      0.647 &      0.711 \\

         1 &      1.059 &      1.168 &          2 &      2.257 &      2.462 &        0.5 &      0.592 &      0.633 \\

         2 &      2.042 &      2.123 &          2 &      2.163 &      2.297 &        0.5 &      0.562 &      0.590 \\

         5 &      5.029 &      5.088 &          2 &      2.097 &      2.184 &        0.5 &      0.541 &      0.559 \\

        10 &     10.024 &     10.073 &          2 &      2.066 &      2.129 &        0.5 &      0.530 &      0.543 \\

       0.5 &      0.564 &      0.643 &        0.5 &      0.741 &      0.917 &        0.8 &      0.817 &      0.836 \\

         1 &      1.035 &      1.081 &        0.5 &      0.612 &      0.686 &        0.8 &      0.805 &      0.816 \\

         2 &      2.014 &      2.036 &        0.5 &      0.553 &      0.592 &        0.8 &      0.804 &      0.811 \\

         5 &      5.001 &      5.005 &        0.5 &      0.517 &      0.533 &        0.8 &      0.803 &      0.807 \\

        10 &      9.999 &      9.999 &        0.5 &      0.507 &      0.513 &        0.8 &      0.802 &      0.804 \\

       0.5 &      0.638 &      0.835 &          2 &      2.348 &      2.586 &        0.8 &      0.835 &      0.860 \\

         1 &      1.121 &      1.278 &          2 &      2.194 &      2.303 &        0.8 &      0.819 &      0.835 \\

         2 &      2.111 &      2.249 &          2 &      2.137 &      2.212 &        0.8 &      0.813 &      0.823 \\

         5 &      5.108 &      5.236 &          2 &      2.101 &      2.165 &        0.8 &      0.811 &      0.818 \\

        10 &     10.118 &      10.250      &          2 &      2.081 &    2.139        &        0.8 &      0.810 &       0.816     \\
\hline
\end{tabular}

\caption{\label{nb2-bias}Limiting DTMLE for   MVN copula-based models  with NB2($\mu;\gamma$) margins. The truncation point is chosen to exceed 0.999 for total probabilities.}
\end{center}
\end{table}

To sum up,  the DT approximation is worse with more discretization (larger individual probabilities). The result should not be in terms of  higher mean values as in \cite{Kazianka2013}, but in terms of the size of the discrete probabilities. For higher mean values and moderate overdispersion, the probability of any given event will decrease, and as a result,
one would be linearly approximating `smaller steps' in the cdf.

After  evaluating the adequacy of the simulated  and surrogate  log-likelihood on finding  the peak (MLE), we evaluate if the curvature (Hessian) is also correct for the cases where the MSLE and DTMLE are correct. To check this, we also computed the negative inverse Hessian $H$ of the limit of the surrogate log-likelihood  in (\ref{limitlik}) and the  simulated log-likelihood in (\ref{limitslik}); because these are limits as $n\to\infty$ of $n^{-1}$ times the log-likelihood, $H$ is the inverse Fisher information, or equivalently, the
covariance matrix for sample size $n$ is approximately $n^{-1}H$.  For a comparison, we have also calculated the Hessian at the limit for the standard MLE. For simpler comparisons, we convert to standard errors (SE), say for a sample size of $n = 100$ (that is, square roots of the
diagonals of the above matrices divided by $n$). Some representative results are given in Table \ref{pois-se} for MVN copula-based models with Poisson margins, with the MSLE results omitted because they were again identical with MLE up to three or four decimal places.
The results in Table \ref{pois-se} show that the surrogate likelihood  method slightly underestimates the SEs and  underestimation of the curvature increases as  the dimension $d$ and/or the latent correlation $\rho$ increases when the DTMLE are correct.

\begin{table}[!h]
\begin{center}

\begin{tabular}{cc|cccc|cccc}
\hline
    $\mu$ &    $\rho$ &                   \multicolumn{ 4}{c|}{SE($\mu$)} &                  \multicolumn{ 4}{c}{SE($\rho$)} \\

           &            & \multicolumn{ 2}{c}{$d=2$} & \multicolumn{ 2}{c|}{$d=3$} & \multicolumn{ 2}{c}{$d=2$} & \multicolumn{ 2}{c}{$d=3$} \\
           &            &         ML &        SUL &         ML &        DT &         ML &        DT &         ML &        DT \\
\hline
       0.5 &        0.2 &      0.054 &      0.055 &      0.046 &      0.050 &      0.135 &      0.133 &      0.087 &      0.088 \\

         1 &        0.2 &      0.076 &      0.077 &      0.067 &      0.067 &      0.112 &      0.120 &      0.073 &      0.082 \\

         2 &        0.2 &      0.109 &      0.109 &      0.096 &      0.096 &      0.101 &      0.105 &      0.067 &      0.071 \\

         5 &        0.2 &      0.173 &      0.173 &      0.151 &      0.152 &      0.096 &      0.097 &      0.064 &      0.065 \\

        10 &        0.2 &      0.244 &      0.245 &      0.217 &      0.215 &      0.095 &      0.096 &      0.063 &      0.063 \\

       0.5 &        0.5 &      0.059 &      0.059 &      0.055 &      0.055 &      0.105 &      0.070 &      0.074 &      0.047 \\

         1 &        0.5 &      0.085 &      0.084 &      0.079 &      0.077 &      0.086 &      0.071 &      0.061 &      0.049 \\

         2 &        0.5 &      0.121 &      0.120 &      0.114 &      0.111 &      0.075 &      0.070 &      0.053 &      0.049 \\

         5 &        0.5 &      0.193 &      0.192 &      0.181 &      0.180 &      0.070 &      0.068 &      0.049 &      0.048 \\

        10 &        0.5 &      0.273 &      0.273 &      0.257 &      0.256 &      0.068 &      0.068 &      0.048 &      0.047 \\

       0.5 &        0.8 &      0.065 &      0.063 &      0.063 &      0.059 &      0.055 &      0.032 &      0.040 &      0.022 \\

         1 &        0.8 &      0.093 &      0.089 &      0.091 &      0.084 &      0.043 &      0.032 &      0.031 &      0.023 \\

         2 &        0.8 &      0.133 &      0.129 &      0.130 &      0.123 &      0.036 &      0.031 &      0.025 &      0.022 \\

         5 &        0.8 &      0.211 &      0.209 &      0.207 &      0.202 &      0.031 &      0.029 &      0.022 &      0.021 \\

        10 &        0.8 &      0.317 &      0.298 &      0.294 &      0.290 &      0.031 &      0.029 &      0.021 &      0.020 \\
\hline
\end{tabular}
\caption{\label{pois-se}Standard errors (SE) of the limiting DTMLE and MLE for MVN copula-based models with Poisson($\mu$) margins. The truncation point is chosen to exceed 0.999 for total probabilities.}
\end{center}
\end{table}

\section{\label{sec-sim}Small-sample efficiency based on simulation studies} 
In this section we study the small-sample efficiency of the likelihood  estimation methods in the case of discrete data with high-dimensional dependence.
In particular we concentrate on modelling spatially aggregated (areal) binary and count  data.  

Let $G = (V,E)$ be the underlying graph, where $V = \{1,2,...,d\}$ are the vertices and $E\subset V\times V$ are the edges of $G$. Each vertex of G corresponds to a region over which measurements have been aggregated, and each edge of $G$ represents the spatial adjacency of two such regions. 
Let $w_j$ be the degree of vertex $j$, let $\D = \mbox{diag}(w_1,\ldots,w_d)$, let $\varrho\in[0,1)$, and let $\A = \Bigl[1\bigl\{(i,j)\in E\bigr\}\Bigl]$
be the adjacency matrix of  $G$ with rows and columns labelled by graph vertices, with a 1 or 0 in position $(j,k)$ according to whether $j$ and $k$ are adjacent or not.

We adopt the same dependence model, the  so named CAR copula, proposed by \cite{Hughes2014}.
 This is a MVN copula  constructed by the inversion method in (\ref{inversion-method}). 
In particular,
if $\Phi_d(\cdot;\S)$
is the MVN cdf with covariance matrix 
$$\S=(\s_{jk}: 1\le j\leq k\le d),$$
and $N(0,\s_{jj}^2)$ margins, and $\Phi(\cdot;\s_{jj}^2)$'s are the univariate normal cdfs with variances $\s_{jj}^2$'s,
then the CAR copula is
$$C(u_1,\ldots,u_d;\S)=\Phi_d\Bigl(\Phi^{-1}(u_1;\s_{11}^2),\ldots,\Phi^{-1}(u_d,\s_{dd}^2);\S\Bigr),$$
where $\S=(\D-\varrho \A)^{-1}$  is the inverse of the precision matrix of  the proper
conditionally autoregressive (CAR) model  \citep{Mardia1988,Gelfand&Vounatsou-2003}.
\cite{Hughes2014} considered only Poisson$(\mu)$ margins.
Here we also consider Bernoulli$(\mu)$, NB$1(\mu,\,\ga)$,  and NB$2(\mu,\,\ga)$ parametrization  of the negative binomial distribution to allow for a comprehensive comparison. For the covariates we use the same design, i.e.,  
we chose $p=2, \x_j=(x_{1j},x_{2j},)^\top,$ where $x_{1j}$
and $x_{2j}$ are the coordinates of vertex $j$ and let $\mu$ depend on the covariates, that is $\nu_j=\eta(\mu_j)=\beta_1 x_{1j} + \beta_2 x_{2j}, \,j=1,\ldots,d.$ 
For the regression parameters we use $\beta_1=-\beta_2=2$  for count and binary data, respectively. 
These regression coefficients lead to similar mean values with the ones  used by \cite{Hughes2014}, but the size of the discrete probabilities increases. 
For the link function $\eta$, we took the log link function for Poisson and NB regression, and the logit link function or the probit link function for binary regression. 
Note also that binary and Poisson regression $\gbf$ is null, while for NB1 and NB2 regression $\gamma$ is scalar ($r=1$).

We  randomly generate data on the $15\times 15$ square lattice, where the coordinates of the vertices were restricted to the unit square. We choose to focus on this  lattice size since the Slovenia stomach cancer data \citep{Zadnik&Reich-2006} in the forthcoming section has a similar  dimension. 
We also assume that we have a truly small sample that is just $n = 1$ observation per lattice.
For the CAR copula model and parameter and design selections we derive the estimates via the DT and SL method.

\begin{table}[!h]
\begin{center}
\begin{tabular}{c|cc|cc|cc}
\hline
                                                \multicolumn{ 7}{c}{Logistic regression} \\
\hline
           & \multicolumn{ 2}{c|}{$\b_1=2$} & \multicolumn{ 2}{c|}{$\b_2=-2$} & \multicolumn{ 2}{c}{$\varrho=0.5$} \\

           &         SL &         DT &         SL &         DT &         SL &         DT \\
\hline

 $d$ Bias &       9.35 &    -139.41 &      -6.74 &     140.93 &     -19.21 &     109.39 \\

   $d$ SD &     111.66 &      74.03 &     110.34 &      73.76 &      49.87 &       1.33 \\

 $d$ RMSE &     112.05 &     157.85 &     110.54 &     159.07 &      53.44 &     109.40 \\

 \hline                                                \multicolumn{ 7}{c}{Poisson regression} \\
\hline
           & \multicolumn{ 2}{c|}{$\b_1=2$} & \multicolumn{ 2}{c|}{$\b_2=-2$} & \multicolumn{ 2}{c}{$\varrho=0.5$} \\

           &         SL &         DT &         SL &         DT &         SL &         DT \\
\hline
 $d$ Bias &      -3.12 &      -1.68 &       0.93 &      -3.52 &     -11.69 &      21.13 \\

   $d$ SD &      31.86 &      32.32 &      51.94 &      51.61 &      39.19 &      47.56 \\

 $d$ RMSE &      32.01 &      32.36 &      51.95 &      51.73 &      40.89 &      52.04 \\

\end{tabular}\end{center}
\vspace{-1cm}
\begin{center}

\begin{tabular}{c|cc|cc|cc|cc}\hline\multicolumn{ 9}{c}{NB1 regression} \\
\hline & \multicolumn{2}{c|}{$\b_1=2$} & \multicolumn{ 2}{c|}{$\b_2=-2$} & \multicolumn{ 2}{c|}{$\g=2$} & \multicolumn{ 2}{c}{$\varrho=0.5$} \\

           &         SL &         DT &         SL &         DT &         SL &         DT &         SL &         DT 
\\\hline

 $d$ Bias &      -3.21 &      27.75 &      -3.68 &     165.59 &     -12.12 &     684.34 &     -16.82 &      61.67 \\

$d$ Bias* &      -4.58 &      21.15 &      -1.84 &     184.71 &     -17.39 &     714.68 &     -13.17 &     109.74 \\

   $d$ SD &      51.36 &      75.69 &      80.74 &     150.81 &      96.91 &     670.25 &      42.39 &      66.72 \\

 $d$ RMSE &      51.46 &      80.62 &      80.82 &     223.97 &      97.67 &     957.89 &      45.61 &      90.85 \\

$d$ RMSE* &      51.57 &      78.59 &      80.76 &     238.45 &      98.46 &     979.80 &      44.39 &     128.43 \\ \hline \multicolumn{ 9}{c}{NB2 regression} \\
\hline& \multicolumn{ 2}{c|}{$\b_1=2$} & \multicolumn{ 2}{c|}{$\b_2=-2$} & \multicolumn{ 2}{c|}{$\g=2$} & \multicolumn{ 2}{c}{$\varrho=0.5$} \\

           &         SL &         DT &         SL &         DT &         SL &         DT &         SL &         DT 
\\\hline

 $d$ Bias &      -7.91 &     281.34 &      -1.24 &     210.98 &      -7.81 &     166.90 &     -15.20 &      64.98 \\

$d$ Bias* &      -6.60 &     202.64 &       3.00 &     156.28 &     -16.55 &     173.89 &     -11.44 &     111.29 \\

   $d$ SD &      78.00 &     340.96 &      89.80 &     236.34 &      86.27 &     165.86 &      42.53 &      64.34 \\

 $d$ RMSE &      78.40 &     442.05 &      89.81 &     316.81 &      86.62 &     235.30 &      45.16 &      91.45 \\

$d$ RMSE* &      78.28 &     396.63 &      89.85 &     283.34 &      87.84 &     240.30 &      44.04 &     128.55 \\\hline
\end{tabular}\end{center}\caption{\label{sim}The results of the simulation study for the $15\times 15$ lattice and logistic, Poisson, NB1 and NB2 regression for  spatially aggregated (areal) binary and count  data for $\varrho=0.5$.}
\end{table}

\begin{table}[!h]
\begin{center}
\begin{tabular}{c|cc|cc|cc}
\hline
                                                \multicolumn{ 7}{c}{Logistic regression} \\
\hline
           & \multicolumn{ 2}{c|}{$\b_1=2$} & \multicolumn{ 2}{c|}{$\b_2=-2$} & \multicolumn{ 2}{c}{$\varrho=0.8$} \\

           &         SL &         DT &         SL &         DT &         SL &         DT \\
\hline

    $d$ Bias & 15.39 & -142.78 & -12.62 & 144.13 & -18.56 & 43.09 \\
   
    $d$ SD & 148.03 & 95.82 & 144.56 & 94.12 & 36.94 & 0.80 \\
    $d$ RMSE & 148.83 & 171.96 & 145.11 & 172.14 & 41.35 & 43.10 \\
 \hline                                                \multicolumn{ 7}{c}{Poisson regression} \\
\hline
           & \multicolumn{ 2}{c|}{$\b_1=2$} & \multicolumn{ 2}{c|}{$\b_2=-2$} & \multicolumn{ 2}{c}{$\varrho=0.8$} \\

           &         SL &         DT &         SL &         DT &         SL &         DT \\
\hline
    $d$ Bias & -4.83 & -1.47 & -0.75 & 7.61  & -9.57 & 18.49 \\
   
    $d$ SD & 43.11 & 44.69 & 69.47 & 66.00    & 25.56 & 19.36 \\
    $d$ RMSE & 43.38 & 44.72 & 69.47 & 66.43 & 27.30  & 26.77 \\
   \end{tabular}\end{center}
\vspace{-1cm}
\begin{center}

\begin{tabular}{c|cc|cc|cc|cc}\hline\multicolumn{ 9}{c}{NB1 regression} \\
\hline& \multicolumn{ 2}{c|}{$\b_1=2$} & \multicolumn{ 2}{c|}{$\b_2=-2$} & \multicolumn{ 2}{c|}{$\g=2$} & \multicolumn{ 2}{c}{$\varrho=0.8$} \\
          &         SL &         DT &         SL &         DT &         SL &         DT &         SL &         DT 
\\\hline

    $d$ Bias & -5.02 & 38.84 & -6.88 & 273.67 & -21.33 & 1036.08 & -15.62 & 40.56 \\
   
    $d$ SD & 68.67 & 91.15 & 104.23 & 104.47 & 100.91 & 537.88 & 28.68 & 14.50 \\
    $d$ RMSE & 68.86 & 99.08 & 104.46 & 292.93 & 103.14 & 1167.38 & 32.66 & 43.08 \\

 \hline \multicolumn{ 9}{c}{NB2 regression} \\
\hline& \multicolumn{ 2}{c|}{$\b_1=2$} & \multicolumn{ 2}{c|}{$\b_2=-2$} & \multicolumn{ 2}{c|}{$\g=2$} & \multicolumn{ 2}{c}{$\varrho=0.8$} \\
          &         SL &         DT &         SL &         DT &         SL &         DT &         SL &         DT 
\\\hline

    $d$ Bias & -11.43 & 434.98 & -3.26 & 342.80 & -11.13 & 239.90 & -14.17 & 41.24 \\
    
    $d$ SD & 99.66 & 335.59 & 113.30 & 232.01 & 99.61 & 138.89 & 28.75 & 13.91 \\
    $d$ RMSE & 100.31 & 549.39 & 113.34 & 413.93 & 100.23 & 277.21 & 32.05 & 43.53 \\
\hline
\end{tabular}\end{center}\caption{\label{sim2}The results of the simulation study for the $15\times 15$ lattice and logistic, Poisson, NB1 and NB2 regression for  spatially aggregated (areal) binary and count  data for $\varrho=0.8$.}
\end{table}

Representative summaries of findings on the performance of the  approaches are given in Table \ref{sim} and Table \ref{sim2} for $\varrho=0.5$ and $\varrho=0.8$, respectively.     
The tables contain the true parameter values, the bias,
standard deviation (SD), and root mean square error (RMSE) scaled by $d=15^2$ of the
DT and SL  estimates from $10^3$ random samples  generated from the CAR copula and marginal logistic, Poisson, NB1 and NB2 regression.
For NB1 and NB2 regression in Table \ref{sim} the distribution of the DT estimators is quite skewed (e.g.,  $\hat\varrho$ is skewed to the upper bound of the parameter space), thus  we also calculate the sample  median to be more informative. Asterisks
indicate the corresponding biases and RMSEs.  

Conclusions from the values in the tables 
are the following:
\begin{itemize}
\itemsep=0pt
\item The SL method is highly efficient according to the simulated biases and variances.

\item The DT 
method yields estimates that are almost as good as the SL estimates for the regression
parameters when there is less discretization (smaller individual probabilities).
\item The DT method overestimates the univariate marginal parameters  
when there is more discretization (larger individual probabilities). 

\item The efficiency of the DT method is low for the parameter $\varrho$ of the CAR precision matrix.  
The parameter $\varrho$ is substantially overestimated and overestimation decreases as $\varrho$ increases or the individual probabilities decrease.

\end{itemize}

\section{\label{sec-app}Illustrations} 
In this section we illustrate the methods  with spatially aggregated count data. In the first subsection we apply the likelihood estimation methods to  the Slovenia stomach cancer  data \citep{Zadnik&Reich-2006},  also analysed in \cite{Hughes2014}. In this areal dataset, there are  small individual probabilities. In the  second subsection the methods are in contrast applied to the Ohio lung cancer incidence data \citep{xia&carlin1998} for which there apparently  exist  large individual probabilities. As emphasized in the preceding sections, the size of the discrete probabilities can substantially influence  the efficiency of the surrogate likelihood method based on the DT. 
In fact in this section we also calculate/plot some simple descriptive statistics to form as diagnostics for the efficiency of  DT method   for the data on hand. 

Model selection is often based on information criteria such as AIC, BIC, SBC or Generalized AIC in order to include a penalty for the different number of parameters among the models. We adopt one of this criteria, namely the AIC, here, since the NB regression with spatial discrete data has an additional parameter. The discussion below
could also apply to other information criteria. 
By using the DT or SL method,
the AIC is $-2\times$log-likelihood $+2\times$ (\#model parameters)
and a smaller AIC value indicates a better fitting model.

\subsection{The Slovenia data}
In this section we re-analyse the Slovenia stomach cancer incidence data in \cite{Hughes2014}. The Slovenia cancer incidence data consist of municipality-level ($d=194$)  observed deaths from stomach cancer in Slovenia for the period 1995-2001 \citep{Zadnik&Reich-2006} and are provided at the supplementary material in \cite{Hodges2013}.  Number of expected deaths from stomach cancer and municipality specific socio-economic statuses  as determined by Slovenia's Institute of Macroeconomic Analysis and Development are also available. 
The interest of this analysis is to
explain the relationship of the stomach cancer cases as a function of
the  municipality  specific
socio-economic statuses.

\begin{figure}[!h]
\begin{center}
\begin{tabular}{cc}
\multicolumn{2}{c}{
\includegraphics[width=0.5\textwidth]{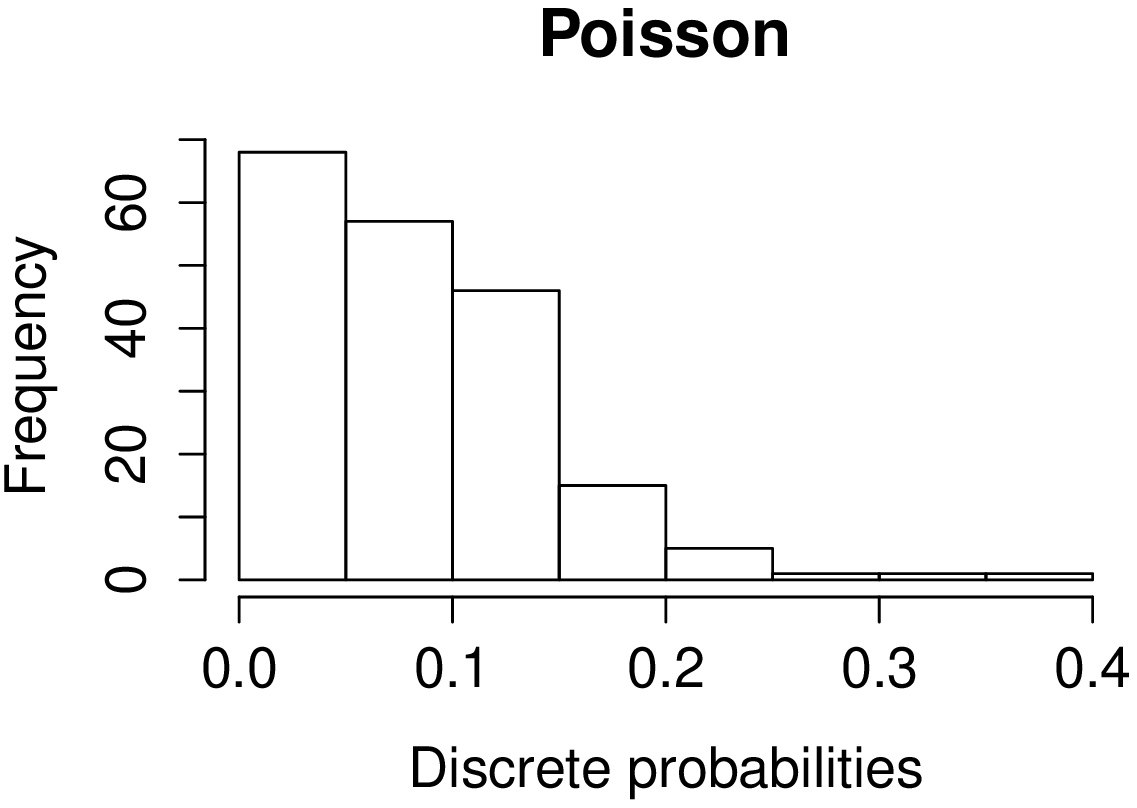}}\\
\includegraphics[width=0.5\textwidth]{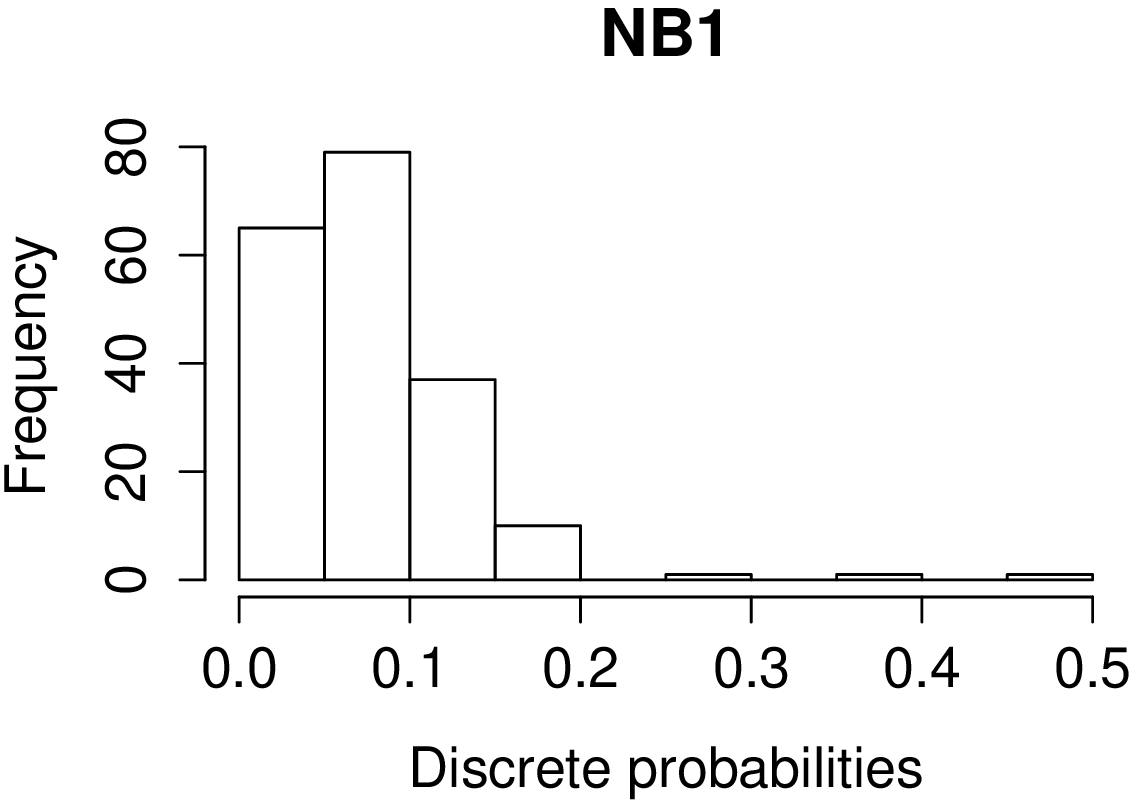} &
\includegraphics[width=0.5\textwidth]{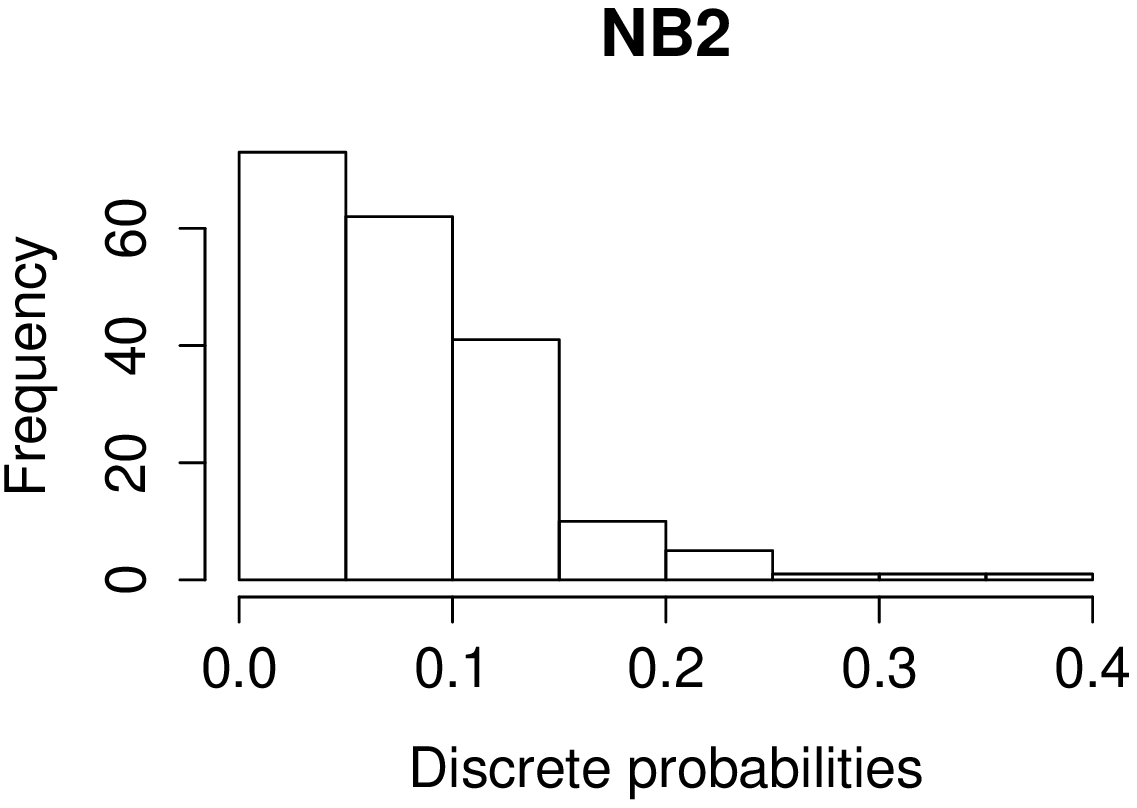}
\end{tabular}
\caption{\label{size}The size of the the discrete probabilities assuming independence for the Slovenia stomach cancer data.}
\end{center}
\end{figure}

The observed stomach cancer cases $y_{j},\,j=1,\ldots,194$ are  counts, so we can assume a marginal Poisson or NB1 or NB2 model with means:
$$\mu_{j}=\mbox{E}_{j}\exp(\beta_0+\beta_1\mbox{SoEc}_{j}),\,j=1,\ldots,194,$$
where $\mbox{E}_{j}$ and $\mbox{SoEc}_{j}$ is the expected number of cases and standardised socio-economic status respectively for municipality $j$.
For a preliminary analysis, we fit the model ignoring the spatial dependence, that is we assume independence. Figure \ref{size} depicts the size of the estimated discrete probabilities for all models under the independence assumption.  As revealed there are some small individual probabilities, suggesting that the surrogate likelihood method based on the DT might  be reliable.

Then, we fit the discrete  CAR  copula models  performing estimation via the DT and SL methods.
Table \ref{slovenia} gives the
estimated parameters along with  the AIC values.  The best fit in terms of the penalized log-likelihood principle (AIC) is based on the MVN copula with an NB1 margin, where there is an improvement over the  MVN copula with a Poisson. The AIC values show that NB1 regression is marginally better than NB2 regression, and both are far better than Poisson regression \citep{Hughes2014}. 
The estimate of the parameter $\varrho$ of the CAR precision matrix is significantly underestimated under the assumption of a Poisson margin; there is enough improvement to change a value of $\hat\varrho=0.29$ to one of $\hat\varrho=0.44$. 

\begin{table}[!h]
\begin{center}
\begin{tabular}{c|cc|cc|cc}
\hline
           & \multicolumn{ 2}{c|}{Poisson regression} & \multicolumn{ 2}{c|}{NB1 regression} & \multicolumn{ 2}{c}{NB2 regression} \\

           &         DT &         SL &         DT &         SL &         DT &         SL \\
\hline
   $\b_0$ &      0.153 &      0.153 &      0.144 &      0.145 &      0.144 &      0.145 \\

   $\b_1$ &     -0.128 &     -0.128 &     -0.120 &     -0.120 &     -0.098 &     -0.098 \\

     $\g$ &      -      &      -      &      0.893 &      0.891 &      0.047 &      0.047 \\

$\varrho$ &      0.283 &      0.289 &      0.438 &      0.438 &      0.436 &      0.436 \\
\hline
    AIC         & 1150.7 & 1150.4 & 1110.6 & 1110.4 & 1115.2 & 1115.0 \\

\hline
\end{tabular}\caption{\label{slovenia}Estimated parameters and  AIC values using the DT and SL methods for the Slovenia stomach cancer data.} 

\end{center}
\end{table}

\begin{figure}[!h]
\begin{center}
\begin{tabular}{cc}
\includegraphics[width=0.5\textwidth]{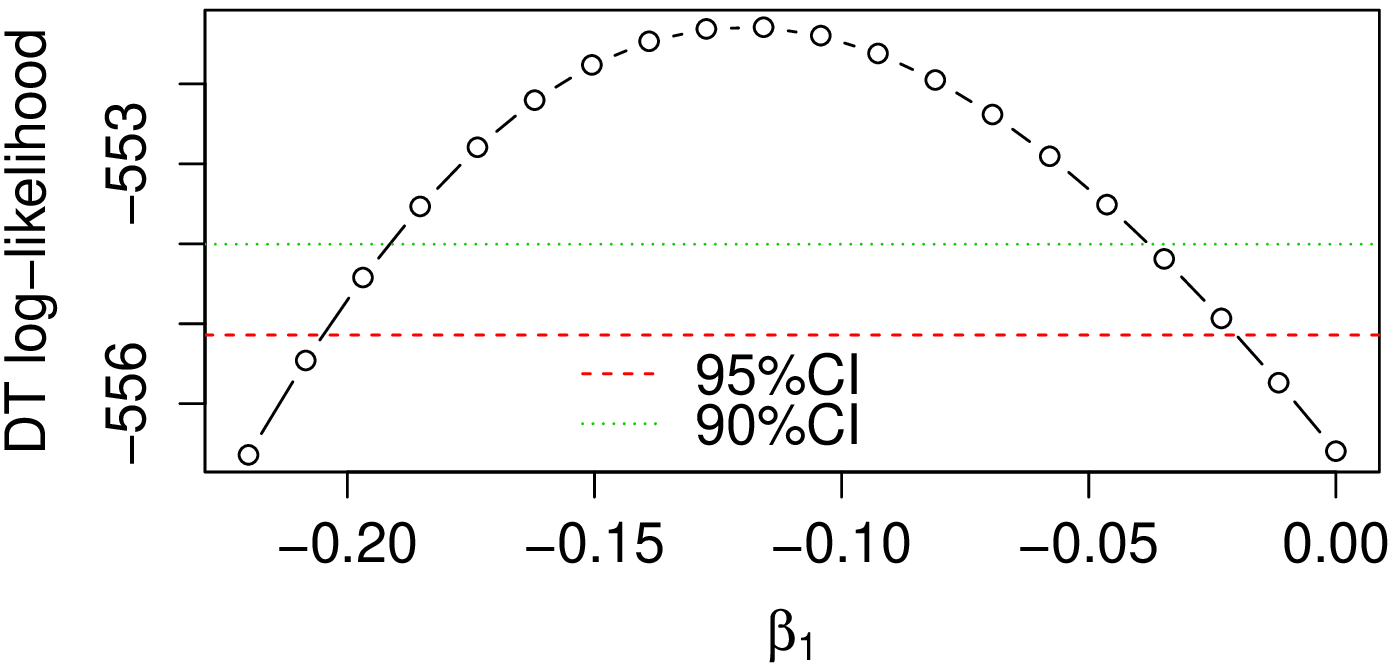}
&
\includegraphics[width=0.5\textwidth]{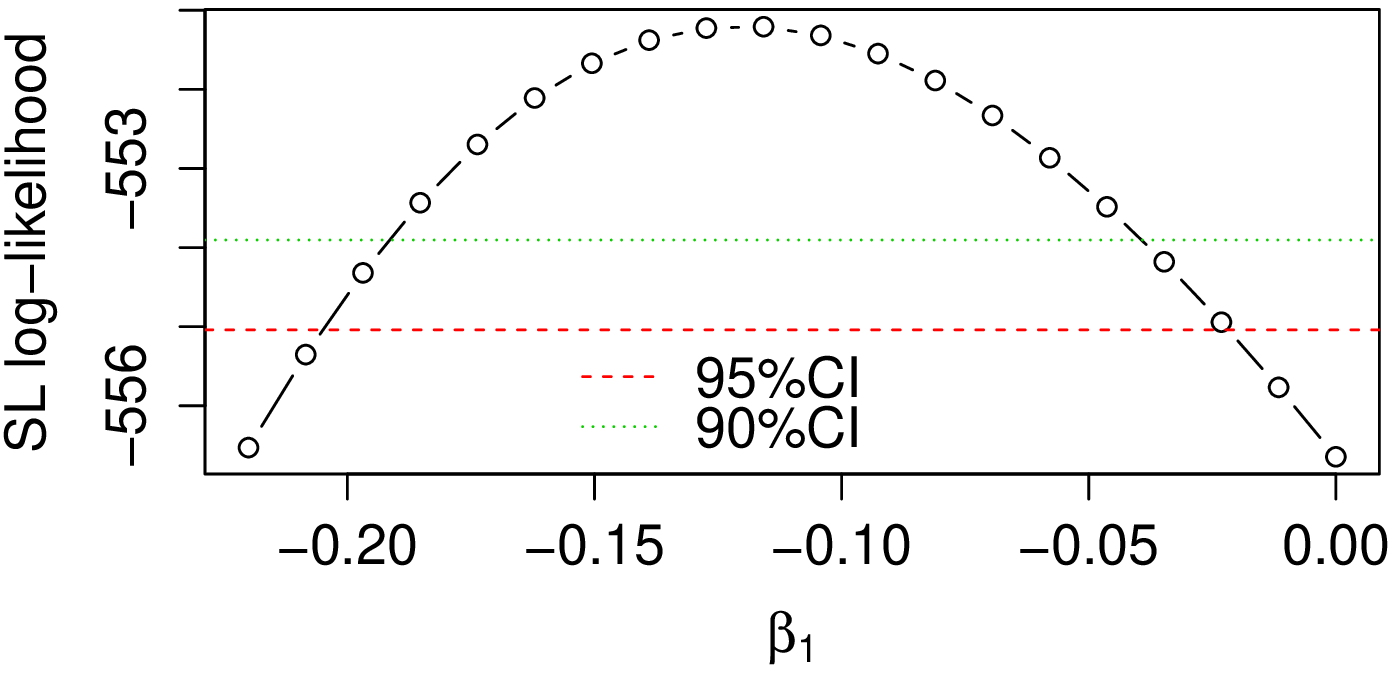}
\end{tabular}
\caption{\label{profileCIs}Profile likelihoods for the SoEc coefficient for the MVN copula with NB1 margins model fitted to the Slovenia stomach cancer data. The dashed and dotted horizontal lines indicate the 95\% and 90\% confidence intervals, respectively.}
\end{center}
\end{figure}

The estimated regression coefficient evidences a negative effect of SoEc  with the number of cases. A further confirmation of the effect between cases and SoEc is given by the log-profile likelihoods of the SoEc coefficient displayed in Figure \ref{profileCIs}. 
The profile CIs show the SoEc is   significantly negatively associated with excess of stomach cancer incidence. Our analysis also implies that the socio-economic status does not account for all of the spatial confounding in the count response since the estimate of  $\varrho$ was as large as $0.44$.

\subsection{The Ohio data}
The Ohio lung cancer data consist of county-level ($d=88$) deaths from lung cancer in Ohio for the period 1968-1988 \citep{xia&carlin1998} and are provided at the supplementary material in \cite{banerjee-etal-2014}. They are stratified on
race (whites vs non-whites) and gender (males vs females); that is there are $n=4$ observations available per county.  The interest of this analysis is to
explain the relationship of the lung cancer cases as a function of
the stratified covariates. Another question of interest was whether  the  difference between males and females was different for whites and non-whites.  
Here we mainly illustrate the methods for the 1975 data only but the analysis can reproduced for the other years as well.
  
\begin{figure}[!h]
\begin{center}
\begin{tabular}{cc}
\multicolumn{2}{c}{
\includegraphics[width=0.5\textwidth]{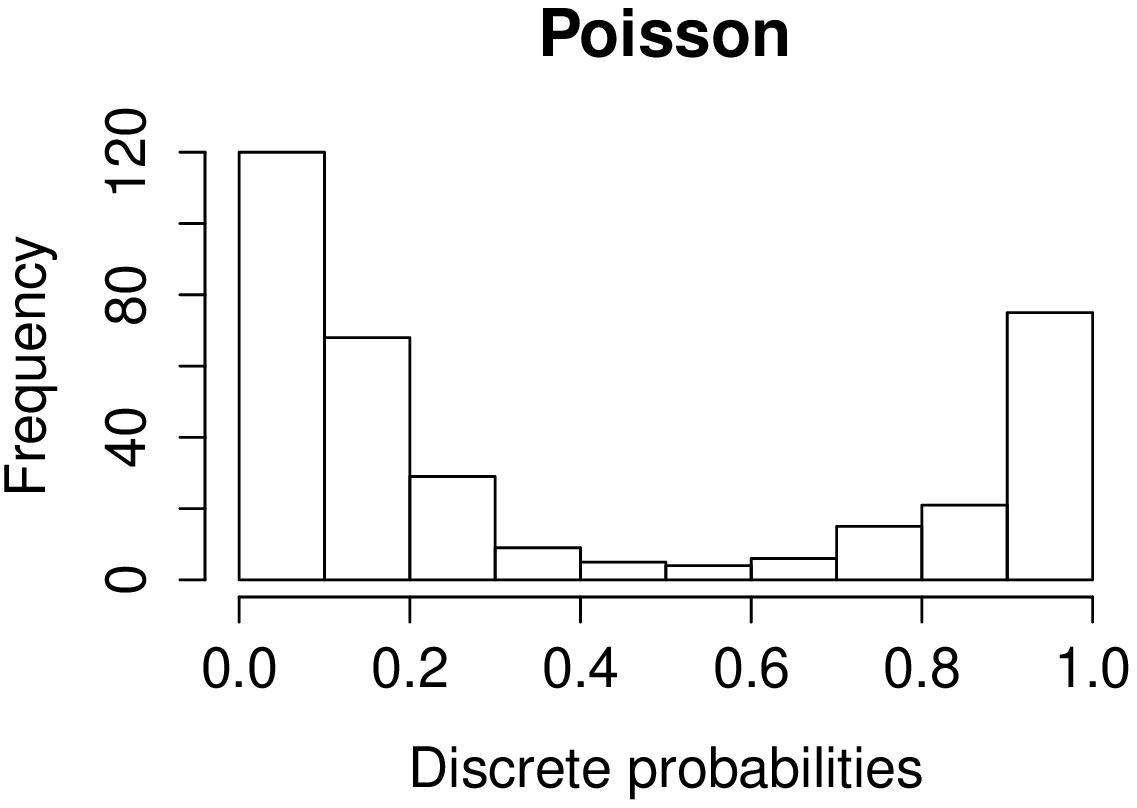}}\\
\includegraphics[width=0.5\textwidth]{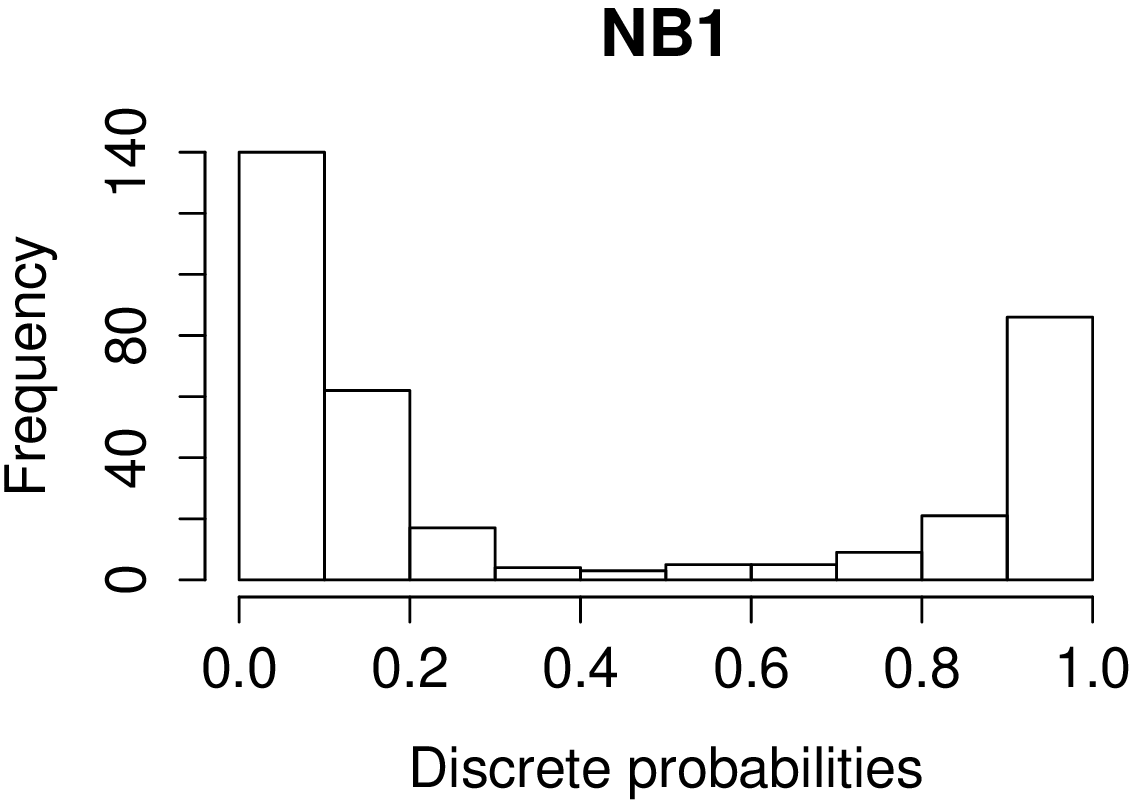} &
\includegraphics[width=0.5\textwidth]{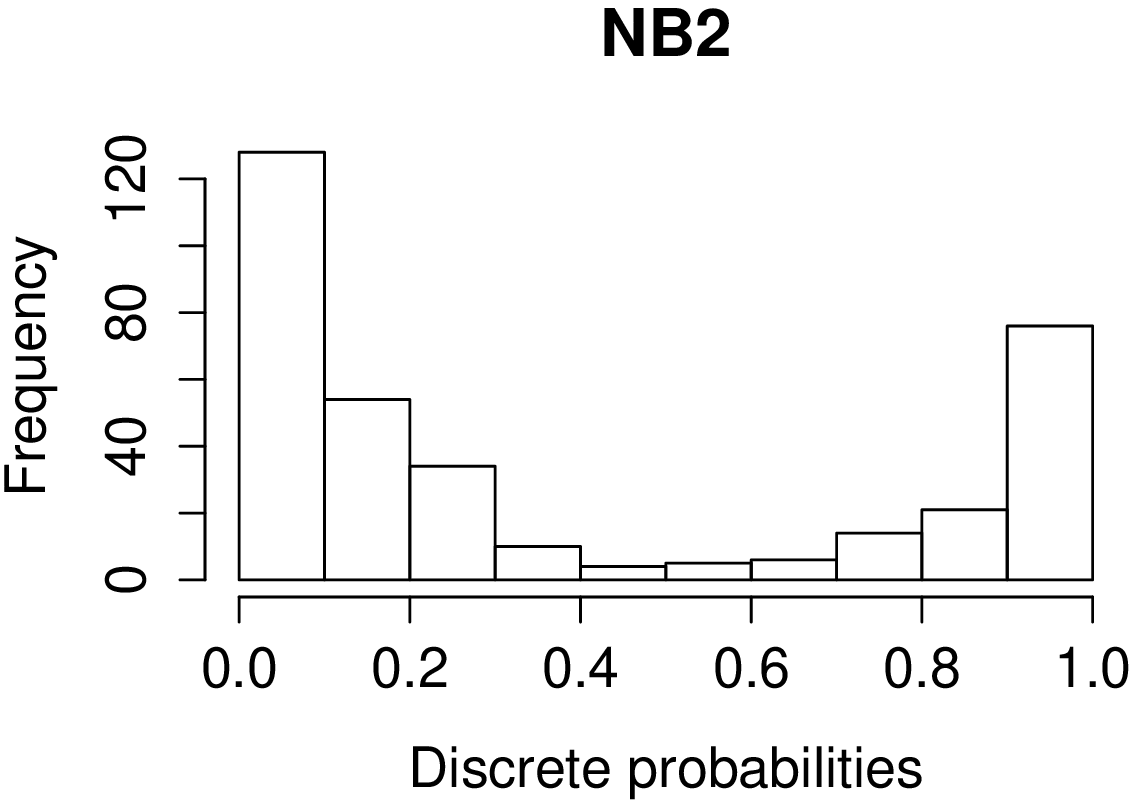}
\end{tabular}
\caption{\label{ohio-size}The size of the the discrete probabilities assuming independence for the Ohio lung  cancer data.}
\end{center}
\end{figure}

The observed lung cancer cases $y_{ij},\,i=1,\ldots,4,\,j=1,\ldots,88$ are  counts, so we can assume a marginal Poisson or NB1 or NB2 model with means:
$$\mu_{ij}=\mbox{E}_{ij}\exp(\beta_0+\beta_1\mbox{Race}_{ij}+\beta_2\mbox{Gender}_{ij}+\beta_3\mbox{Race}_{ij}\times\mbox{Gender}_{ij}),\,i=1,\ldots,4,\,j=1,\ldots,88,$$
where $\mbox{E}_{ij}$  is the expected number of cases. 
For a preliminary analysis, we fit the model ignoring the spatial dependence, that is we assume independence. Figure \ref{ohio-size} depicts the size of the estimated discrete probabilities for all models under the independence assumption.  As revealed there are some large individual probabilities, suggesting that the surrogate likelihood method based on the DT might not be reliable.

\begin{table}[!h]
\begin{center}
\begin{tabular}{c|cccc|cccc}
\hline
           &                                                               \multicolumn{ 8}{c}{Poisson regression} \\
\hline
           &                           \multicolumn{ 4}{c|}{DT} &                           \multicolumn{ 4}{c}{SL} \\

           &       Est. &         SE &       $Z$ & $p$-value &       Est. &         SE &       $Z$ & $p$-value \\
\hline

   $\b_0$ &     -7.331 &      0.029 &   -254.534 &      $<0.001$ &     -7.345 &      0.025 &   -294.174 &      $<0.001$ \\

   $\b_1$ &      0.060 &      0.062 &      0.976 &      0.329 &      0.077 &      0.061 &      1.262 &      0.207 \\

   $\b_2$ &     -1.180 &      0.049 &    -24.148 &      $<0.001$ &     -1.189 &      0.045 &    -26.321 &      $<0.001$ \\

   $\b_3$ &     -0.275 &      0.132 &     -2.076 &      0.038 &     -0.261 &      0.133 &     -1.960 &      0.050 \\

$\varrho$ &      0.540 &      0.166 &      3.247 &      0.001 &      0.385 &      0.147 &      2.626 &      0.009 \\
\hline
    AIC   & \multicolumn{4}{c}{1255.6}  & \multicolumn{4}{c}{1262.1} \\

\hline
           &                                                                   \multicolumn{ 8}{c}{NB1 regression} \\
\hline
           &                           \multicolumn{ 4}{c|}{DT} &                           \multicolumn{ 4}{c}{SL} \\

           &       Est. &         SE &       $Z$ & $p$-value &       Est. &         SE &       $Z$ & $p$-value \\
\hline
   $\b_0$ &     -7.233 &      0.040 &   -182.519 &      $<0.001$ &     -7.343 &      0.028 &   -260.013 &      $<0.001$ \\

   $\b_1$ &      0.037 &      0.079 &      0.469 &      0.639 &      0.085 &      0.076 &      1.107 &      0.268 \\

   $\b_2$ &     -1.079 &      0.078 &    -13.806 &      $<0.001$ &     -1.185 &      0.019 &    -61.184 &      $<0.001$ \\

   $\b_3$ &     -0.242 &      0.154 &     -1.578 &      0.115 &     -0.253 &      0.158 &     -1.602 &      0.109 \\

     $\g$ &      2.452 &      0.488 &      5.028 &      $<0.001$ &      0.596 &      0.227 &      2.621 &      0.009 \\

$\varrho$ &      0.989 &      0.003 &    324.739 &      $<0.001$ &      0.448 &       0.015 &        29.461 &        $<0.001$ \\
\hline
    AIC   & \multicolumn{4}{c}{1197.644}  & \multicolumn{4}{c}{1240.162} \\

\hline
           &                                                                   \multicolumn{ 8}{c}{NB2 regression} \\
\hline
           &                           \multicolumn{ 4}{c|}{DT} &                           \multicolumn{ 4}{c}{SL} \\

           &       Est. &         SE &       $Z$ & $p$-value &       Est. &         SE &       $Z$ & $p$-value \\
\hline
   $\b_0$ &     -7.369 &      0.054 &   -137.178 &      $<0.001$ &     -7.386 &      0.037 &   -202.052 &      $<0.001$ \\

   $\b_1$ &      0.087 &      0.096 &      0.908 &      0.364 &      0.096 &      0.088 &      1.088 &      0.277 \\

   $\b_2$ &     -1.209 &      0.086 &    -14.104 &      $<0.001$ &     -1.237 &      0.066 &    -18.735 &      $<0.001$ \\

   $\b_3$ &     -0.247 &      0.174 &     -1.417 &      0.156 &     -0.219 &      0.166 &     -1.322 &      0.186 \\

     $\g$ &      0.032 &      0.012 &      2.681 &      0.007 &      0.029 &      0.010 &      3.015 &      0.003 \\

$\varrho$ &      0.675 &      0.266 &      2.537 &      0.011 &      0.331 &      0.183 &      1.807 &      0.071 \\
\hline
    AIC   & \multicolumn{4}{c}{1225.601}  & \multicolumn{4}{c}{1232.910} \\
\hline
\end{tabular}
\caption{\label{ohio}Estimated parameters, standard errors (SE) and  AIC values using the DT and SL methods for the Ohio lung  cancer data for the 1975 data.} 
\end{center}
\end{table}

Then, we fit the discrete  CAR  copula models  performing estimation via the DT and SL methods.
Table \ref{ohio} gives the
estimated parameters and their standard errors (SE) along with  the AIC values. 
The SEs of the maximum SL and DT estimates are obtained via the gradients and the Hessian computed numerically during the maximization process. Assuming that the usual regularity conditions \citep{serfling80} for
asymptotic maximum likelihood theory hold for the bivariate model
as well as for its margins we have that the estimates are
asymptotically normal. Therefore we also build and present Wald tests to
statistically judge the effect of any covariate.
The best fit in terms of the penalized log-likelihood principle (AIC) is based on the MVN copula with an NB2 margin, where there is an improvement over the  MVN copula with a Poisson. The AIC values show that NB2 regression is marginally better than NB1 regression, and both are far better than Poisson regression.

\begin{table}[!h]
\begin{center}
\begin{tabular}{c|cccc|cccc}
\hline

           &                                                               \multicolumn{ 8}{c}{Poisson regression} \\
\hline
           &                           \multicolumn{ 4}{c|}{DT} &                           \multicolumn{ 4}{c}{SL} \\

           &       Est. &         SE &       $Z$ & $p$-value &       Est. &         SE &       $Z$ & $p$-value \\
\hline
   $\b_0$ &     -7.429 &      0.024 &   -308.975 &      $<0.001$ &     -7.429 &      0.024 &   -311.736 &      $<0.001$ \\

   $\b_1$ &      0.134 &      0.060 &      2.250 &      0.024 &      0.137 &      0.059 &      2.315 &      0.021 \\

   $\b_2$ &     -1.276 &      0.043 &    -29.854 &      $<0.001$ &     -1.276 &      0.041 &    -31.326 &      $<0.001$ \\

   $\b_3$ &     -0.385 &      0.140 &     -2.745 &      0.006 &     -0.382 &      0.140 &     -2.724 &      0.006 \\

$\varrho$ &      0.144 &      0.191 &      0.752 &      0.452 &      0.148 &      0.181 &      0.816 &      0.415 \\
\hline
    AIC   & \multicolumn{4}{c}{1250.759}  & \multicolumn{4}{c}{1255.702} \\
   
\hline
           &                                                                   \multicolumn{ 8}{c}{NB1 regression} \\
\hline
           &                           \multicolumn{ 4}{c|}{DT} &                           \multicolumn{ 4}{c}{SL} \\

           &       Est. &         SE &       $Z$ & $p$-value &       Est. &         SE &       $Z$ & $p$-value \\
\hline
   $\b_0$ &     -7.257 &      0.042 &   -173.526 &      $<0.001$ &     -7.434 &      0.030 &   -244.893 &      $<0.001$ \\

   $\b_1$ &      0.019 &      0.084 &      0.226 &      0.821 &      0.151 &      0.075 &      2.017 &      0.044 \\

   $\b_2$ &     -1.190 &      0.083 &    -14.372 &      $<0.001$ &     -1.266 &      0.053 &    -23.698 &      $<0.001$ \\

   $\b_3$ &     -0.272 &      0.173 &     -1.576 &      0.115 &     -0.362 &      0.172 &     -2.105 &      0.035 \\

     $\g$ &      1.857 &      0.456 &      4.068 &      $<0.001$ &      0.608 &      0.161 &      3.771 &      $<0.001$ \\

$\varrho$ &      0.976 &      0.012 &     83.868 &      $<0.001$ &      0.145 &      0.259 &      0.558 &      0.577 \\
\hline
    AIC   & \multicolumn{4}{c}{1218.693}  & \multicolumn{4}{c}{1231.642} \\
    
   \hline

           &                                                                   \multicolumn{ 8}{c}{NB2 regression} \\
\hline
           &                           \multicolumn{ 4}{c|}{DT} &                           \multicolumn{ 4}{c}{SL} \\

           &       Est. &         SE &       $Z$ & $p$-value &       Est. &         SE &       $Z$ & $p$-value \\
\hline
   $\b_0$ &     -7.487 &      0.043 &   -173.800 &      $<0.001$ &     -7.491 &      0.039 &   -191.327 &      $<0.001$ \\

   $\b_1$ &      0.156 &      0.092 &      1.696 &      0.090 &      0.169 &      0.090 &      1.886 &      0.059 \\

   $\b_2$ &     -1.251 &      0.070 &    -17.825 &      $<0.001$ &     -1.254 &      0.117 &    -10.710 &      $<0.001$ \\

   $\b_3$ &     -0.365 &      0.179 &     -2.041 &      0.041 &     -0.364 &      0.202 &     -1.804 &      0.071 \\

     $\g$ &      0.035 &      0.011 &      3.075 &      0.002 &      0.033 &      0.010 &      3.186 &      0.001 \\

$\varrho$ &      0.401 &      0.290 &      1.380 &      0.168 &      0.288 &      0.261 &      1.106 &      0.269 \\
\hline
    AIC   & \multicolumn{4}{c}{1200.528}  & \multicolumn{4}{c}{1206.163} \\
    
\hline
\end{tabular}
\caption{\label{ohio2}Estimated parameters, standard errors (SE) and  AIC values using the DT and SL methods for the Ohio lung  cancer data for the 1974 data.}
\end{center}
\end{table}

Because there is much discretization in the data, the dispersion parameter $\g$ and  parameter $\varrho$ of the CAR precision matrix are (substantially under the assumption of an NB1 margin) over-estimated. This was expected for overdispersed count data as shown in the studies of the  properties of the surrogate likelihood estimates in Section \ref{sec-asym} and Section \ref{sec-sim}. 
 
In particular for areal data applications an  interesting advantage of the CAR copula modelling over the classic areal GLM modelling is the ability to recover $\varrho$; see \cite{Hughes2014} and the references therein. So an efficient  estimation of $\varrho$ is highly desirable in this context; hence   the DT method performs poorly in this example.

Based on our analysis, the SEs show the gender effect to be highly significant, and the gender by race interaction insignificant. However, for the DT analysis  with Poisson regression,  the gender by race interaction is statistically significant. Generally speaking, this implies that misspecifying  the model or using the surrogate likelihood method based on the DT could lead to invalid conclusions.

The latter was actually the case for the analysis of the  data for year 1974 in Table \ref{ohio2}.   The DT method resulted a statical  significant ($p$-value=0.041) rather than the true  marginal  statistical significant ($p$-value=0.071) gender by race interaction  under the best model (NB2) via the penalized likelihood principle (AIC).

\section{\label{sec-disc}Discussion}
In this paper we have studied high-dimensional MVN copula models with discrete margins for analysing spatial discrete response data.  We discussed simulated and surrogate  (based on the DT) likelihood estimation methods.
For the binary, Poisson, and negative binomial regression models with the MVN/CAR copula, we have shown that the surrogate likelihood  method based on the DT leads to substantial  upward bias for the estimates  of the latent correlation/parameter of the precision matrix of the CAR model and the univariate marginal parameters  when there is  more discretization, that is large individual discrete probabilities.

We have shown that the  SL method proposed by \cite{nikoloulopoulos13b}, is  highly  efficient for a high-dimensional discrete response up to dimension $d=15^2$. Although there is an issue of computational burden as the dimension and the sample size increase,  this will subside, as computing technology is advancing rapidly. Any comparison of the methods in terms of  computing time is a digression. It is obvious that the DT method is much faster then the SL method, since  a numerically more difficult high-dimensional MVN rectangle probability calculation is replaced with a much simpler computationally MVN density value.
However, theoretically there are still problems for large individual probabilities, since the DT approximation  of `large steps' in the cdf is poor. In fact, we novelty propose simple diagnostics (descriptive statistics such as a histogram) to judge if the DT method is reliable and reduce by its use the computational burden when it is possible (i.e., for small individual probabilities).

It is worth mentioning that the range of possible applications of these tools goes beyond biometric/disease/health data and is of interest also in other fields such as hydrometeorology. A typical example are the binary vectors describing the rainfall occurrence at multiple sites, or the occurrence of simultaneous exceedance of given threshold values in extreme value analysis of floods, droughts, and storms over specified areas. Note also in passing the DT method deteriorates 
for such (binary) response data. 

Finally, the results will be similar for other structured latent correlation structures such as the Mat\'ern isotropic structure used for example in  \cite{madsen09} and \cite{Kazianka2013}.  As previously  emphasized the idea of the DT transform is to replace a numerically more difficult MVN rectangle probability calculation with a simpler MVN density value, and hence it is discrete responses that matter and not the type of the structured correlation structure.

\end{document}